\documentclass[prl,twocolumn,aps,showpacs]{revtex4}

\usepackage{graphicx}
\usepackage{bm}
\usepackage{amsmath}

\begin{document}

\title{Chiral Ladders and the Edges of Chern Insulators}

\author{Dario H\"{u}gel and Bel\'{e}n Paredes}

\affiliation{Instituto de F\'{i}sica Te\'{o}rica CSIC/UAM \\C/Nicol\'{a}s Cabrera, 13-15
Cantoblanco, 28049 Madrid, Spain}

\date{\today}

\begin{abstract} 
The realization and detection of topological phases with ultracold atomic gases is at the frontier of current theoretical and experimental research.
Here, we identify cold atoms in optical ladders subjected to synthetic magnetic fields as readily realizable 
bridges between one-dimensional spin-orbit (time reversal) topological insulators and two-dimensional  Chern insulators.
We reveal three instances of their promising potential:
i) they realize spin-orbit coupling, with the left-right leg degree of freedom playing the role of an effective spin, ii) their energy bands and eigenstates exactly reproduce the topological chiral edge modes of two-dimensional Chern insulators, and iii) they can be tailored to realize a topological phase transition from a trivial to a topological insulating phase. We propose realistic schemes to observe the chiral and topological properties of ladder systems with current optical 
lattice-based experiments. Our findings open a door to the exploration of the physics of the edges of Chern insulators and to the realization of spin-orbit coupling and topological superfluid phases with ultracold atomic gases.
\end{abstract}
\textbf{}

\maketitle

Realizing topological matter in ultracold atomic systems is an ongoing challenge,  with promising novel perspectives for the investigation of topology in many-body quantum systems \cite{Dalibard2011}.
A current focus of attention is the realization of topological insulating phases \cite{Hasan2010,Zhang2011}, which exhibit remarkable transport properties: they are insulating in the bulk, but have protected conducting states on their edge or surface.
In solid-state systems, topological insulators have been realized both in quantum Hall devices \cite{Hasan2010}, where the presence of an external magnetic field leads to quantization of the electronic conductivity, and in materials with spin-orbit coupling \cite{Hasan2010,Zhang2011}, in which the combination of spin-orbit interaction and time reversal symmetry gives rise to spin and momentum locking.
For ultracold atomic gases, the engineering of synthetic gauge potentials \cite{Spielman2009,Aidelsburger2011,Spielman2012}, on the one hand, and of spin-orbit 
couplings \cite{Spielman2011,Zwierlein}, on the other hand, opens important paths for the exploration of topological phases in novel, unprecedentedly clean and well controlled environments \cite{Bloch2008}. 
By exploiting laser induced gauge potentials \cite{Jaksch2003, Gerbier2010, Struck2012}, large uniform magnetic fields could be achieved in near future optical-lattice based experiments \cite{Aidelsburger2011}, allowing to explore the Hofstadter butterfly model \cite{Hofstadter76}. In this paradigmatic example of a Chern insulator, the interplay between the magnetic field and the periodic potential leads to the formation of energy bands characterized by non-trivial topological invariants, the Chern numbers \cite{Thouless, Kohmoto}. It is a challenge to develop schemes to unambiguously characterize the topological character of atomic Chern insulating phases \cite{Umucalilar2008, Goldman2010, Sinova2010, DasSarma2010, Alba2011, Satija2011, Cooper2012}. Especially promising is the possibility of detecting the emergence of edge modes with well defined chirality \cite{DasSarma2010, Goldman2012, Goldman2013a, Goldman2013b} in which current propagates at the boundaries of the system in a topologically protected manner \cite{Hatsugai, Zhang2006}.
Additionally, using suitably arranged lasers that couple different internal states, spin-orbit coupling has been realized recently in bosonic \cite{Spielman2011} and fermionic \cite{Zwierlein} atomic gases. The closely detuned laser beams employed lead, however, to large photon scattering and heating rates. It is a challenge to develop realistic schemes for the realization of spin-orbit coupling that do not suffer from this limitation. Particularly exciting is the possibility of inducing $p$-wave pairing \cite{Zhang2011} in such spin-orbit coupled atomic systems, paving the way towards the observation of Majorana fermions with non-Abelian braiding statistics \cite{Leijnse2012}.


\begin{figure}[t]
\includegraphics[width=\linewidth]{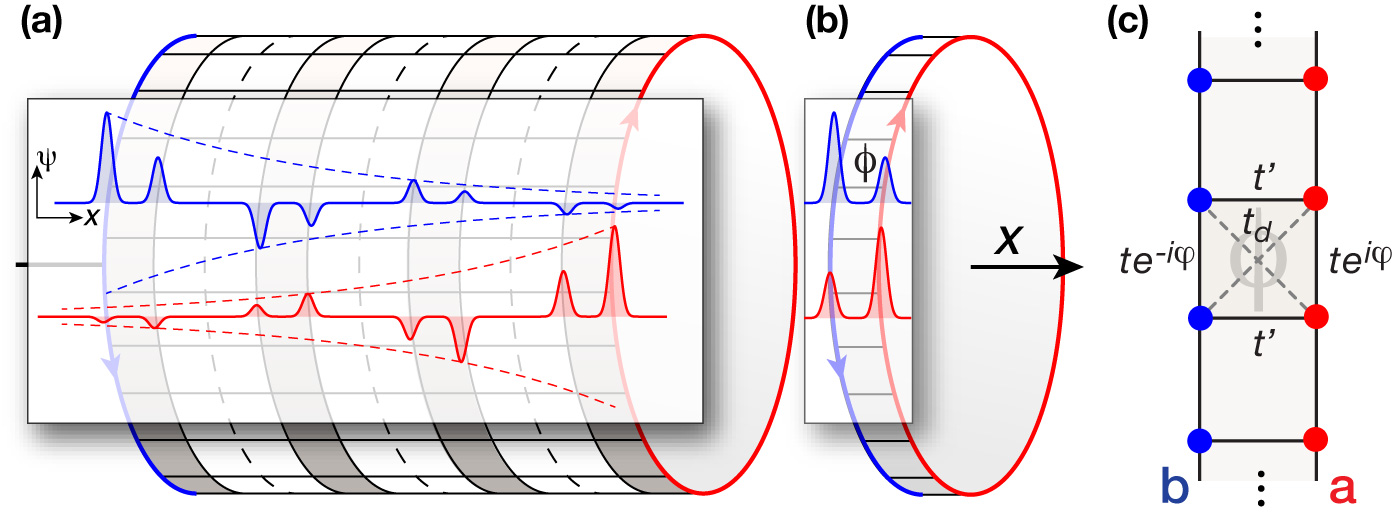}
\caption{\textbf{Ladders and the edges of the Hofstadter model}. 
({\bf a}) Schematic representation of the chiral edge modes of the two-dimensional Hofstadter model. When the system is reduced to a ladder ({\bf b}), the physics of the chiral edge modes remains invariant. 
({\bf c}) Ladder tunneling couplings within the chosen gauge. The unit magnetic cell is a single plaquette, and left and right tunneling amplitudes are complex conjugate of each other.
}
\label{Cylinder-Ladder-Scheme}
\end{figure}
In this work we identify cold atoms in optical ladders subjected to magnetic fields as promising systems that simultaneously realize spin-orbit coupling, and reproduce the physics of the chiral edges of two-dimensional Chern insulators. Ladder systems are readily realized in optical-lattice based experiments that combine effective magnetic fields with superlattice structures \cite{Aidelsburger2011}. Here, we propose realistic schemes in which their chiral and topological properties could be revealed.
In the ladder, the left-right leg degree of freedom plays the role of an effective spin, which couples to the momentum of the particle through the magnetic field. This effective spin-orbit interaction locks spin and momentum, leading to Kramers degenerate pairs \cite{Kramers} of eigenstates with well defined chirality: atoms in the right leg move upwards, whereas atoms in the left leg move downwards.
Unveiling an interesting holographic connection, we show that the energy bands and eigenstates of the ladder system exactly reproduce those characterizing the topological chiral edge modes of the two-dimensional Hofstadter model. This correspondence, which states that shrinking the two-dimensional butterfly model into a strip of plaquettes leaves the physics at the edges invariant, is a manifestation of the topological invariance of the "parent" two-dimensional Chern insulator. 
Moreover, as another instance of their promising potential, we show that ladder systems can be tailored to realize a topological one-dimensional phase transition from a trivial to a topological insulating phase. When diagonal coupling is added to a square geometry, the spin textures characterizing the bands acquire a non-zero winding number. By tuning the ladder parameters, the symmetry class of the topological insulating phase can be changed from the BDI class, the one of polyacetylene, to the chiral unitary (AIII) class \cite{Schnyder}. Polyacetylene-like transitions have been recently observed in a dimerized optical lattice \cite{Atala2012}. They have been also predicted to occur in two-leg ladders with unequal parity orbitals \cite{Liu2013}. 

{\em The ladder}. 
We consider a system of non-interacting particles in a two-leg ladder geometry, subjected to a magnetic flux $\phi$ per plaquette.
We choose a Landau gauge, for which the Hamiltonian is translational invariant in the leg-direction
and the magnetic cell corresponds to a single plaquette [see Fig.\,\ref{Cylinder-Ladder-Scheme}(c)]:
\begin{equation} 
H = -t  \underset{\ell}{\sum}\left(e^{i\varphi}a^{\dagger}_{\ell+1}a^{}_{\ell}+e^{-i\varphi} b^{\dagger}_{\ell+1}b^{}_{\ell} \right)
-t'\underset{\ell}{\sum}a^{\dagger}_{\ell}b^{}_{\ell} \;+ \text{h.c.}
\end{equation}
Here, the operator $a^{}_{\ell}$ ($b^{}_{\ell}$) annihilates a particle at site $\ell$ in the right (left) leg. The hopping amplitude between neighbouring sites in the right (left) leg is $te^{i\varphi}$ ($te^{-i\varphi}$), with $\varphi=\phi/2$, and $t'$ is the tunneling amplitude between legs.
Written in momentum space the Hamiltonian takes the form:
\begin{equation}  
H = 
-2t \;\underset{k}{\sum}\mathbf{c}^{\dagger}_{k}\;\mathcal{H}(k)\;\mathbf{c}^{}_{k},
\end{equation}
with
$\mathbf{c}^{\dagger}_{k}=[a^{\dagger}_{k},b^{\dagger}_{k}]$, 
$a^{\dagger}_{k}(b^{\dagger}_{k})=\frac{1}{\sqrt{L}}\underset{\ell}{\sum}e^{ik\ell}a^{\dagger}_{\ell}(b^{\dagger}_{\ell})$,
and $k=\frac{2\pi}{L}n$, where $n$ is an integer and $L$ is the number of ladder rungs.
The Hamiltonian matrix is 
\begin{equation} 
\mathcal{H}(k)=\varepsilon_0(k){1}+\xi\sigma_x+\sin\varphi\sin k \;\sigma_z,
\label{HamiltonianMomentum}
\end{equation}
where
$\varepsilon_0(k)=\cos\varphi\cos{k}$, 
$\xi=t'/2t$,  and $\sigma_x,\sigma_z$ 
are Pauli matrices.
The system can be thought of as an effective system of spin-$\frac{1}{2}$ particles, where the left-right leg degree of freedom plays the role of an effective spin.
The $\sigma_z$ term in the Hamiltonian is an effective {\em spin-orbit coupling}, which results from the non-zero magnetic flux piercing the ladder. This term leads to spin-momentum locking: spin up (down) particles minimize their energy by having positive (negative) momentum $k$. The ${\sigma}_x$ term is an effective magnetic field in the $x$-direction, which results from tunneling between the ladder legs.
The Hamiltonian is time reversal invariant.  Though it is not real, complex conjugation together with the unitary transformation
$U=\sigma_{x}$, which reverts the particle spin, leaves the Hamiltonian invariant: ${\sigma}_{x}\mathcal{H}^{*}(-k){\sigma}_{x}=\mathcal{H}(k)$.


\begin{figure}[]
\includegraphics[width=\linewidth]{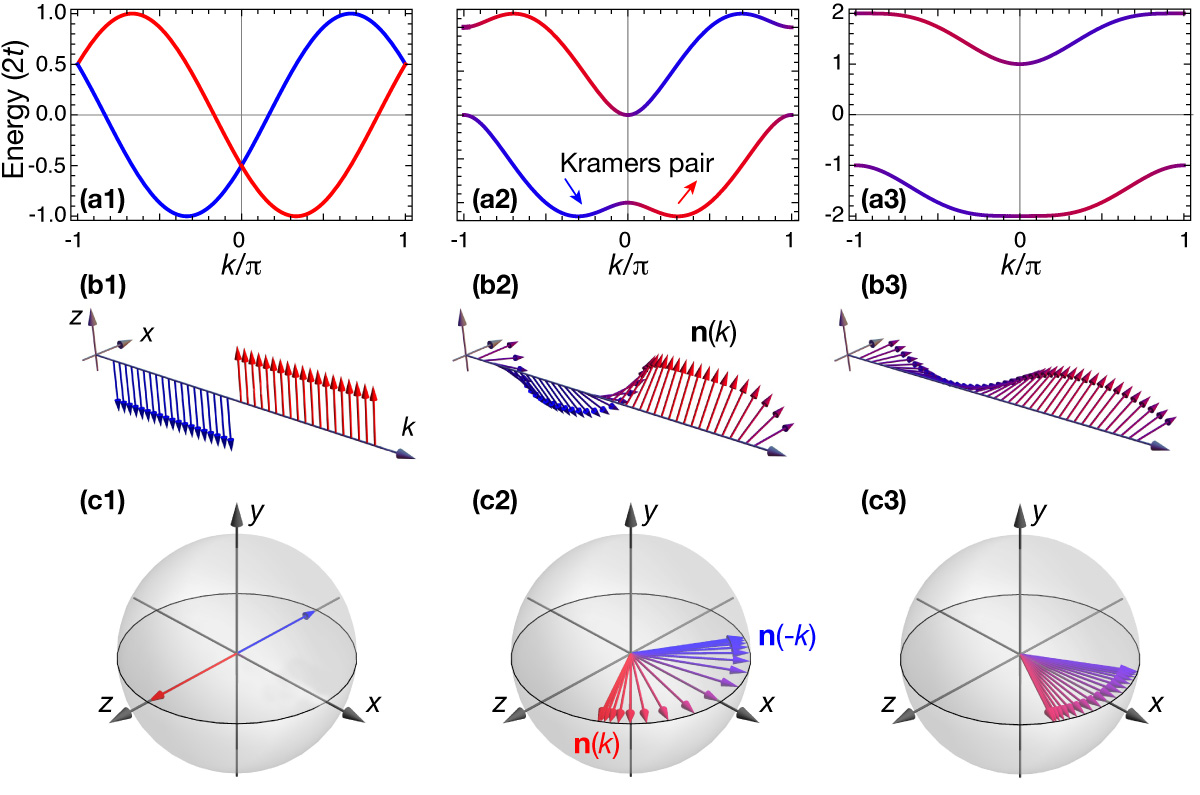}
\caption{\textbf{Ladder energy bands and eigenstates}. Energy bands for flux $\phi=2\pi/3$ and different values of the tunneling between legs, ({\bf a1}) $\xi=0$, ({\bf a2}) $\xi=0.5$, and ({\bf a3}) $\xi=1.5$. The corresponding eigenstates for the lowest band are plotted as Bloch vectors $\hat{n}(k)$ in the first Brillouin zone ({\bf b1},{\bf b2},{\bf b3}) and in the Bloch sphere 
({\bf c1},{\bf c2},{\bf c3}). The vector color represents the average of the spin $z$-component, with red (blue) denoting up (down) states. As the interleg tunneling increases, the arch described by the Bloch vector decreases and the two minima ({\bf a2}) are merged into a single one at momentum zero ({\bf a3}).
}
\label{BlochSpheres}
\end{figure}


\begin{figure}[]
\includegraphics[width=\linewidth]{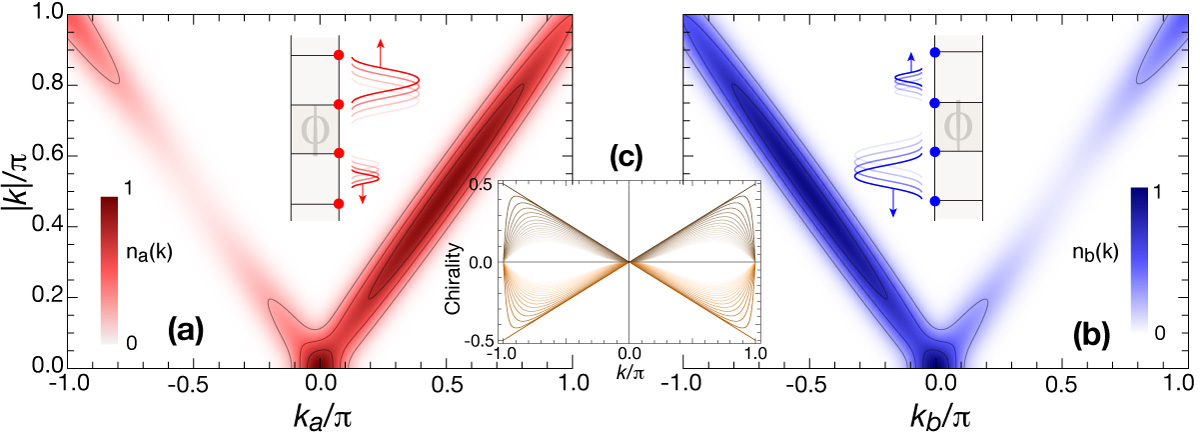}
\caption{\textbf{Spin and momentum locking}. Average momentum density $n_a(k)=\langle a^\dagger_ka^{}_k\rangle$ and $n_b(k)=\langle b^\dagger_kb^{}_k\rangle$ of left ({\bf a}) and right ({\bf b}) components of Kramers pairs within the lowest band. On the average, right (left) particles have positive (negative) momentum and move upwards (downwards). ({\bf c}) Chirality of states within the lowest band (dark brown) and highest band (light brown) for different values of the  interleg tunneling ranging from $\xi=0$ to $\xi=0.5$.
}
\label{SpinMomentum}
\end{figure}

Two energy bands form [Fig.\,\ref{BlochSpheres}(a)], characterized by the energies:
\begin{equation}
E_{\pm}(k)=-\varepsilon_0(k)\pm\sqrt{\rho(k)},
\end{equation}
where $\rho(k)=\xi^2+\sin^2\varphi\sin^2 k$.
The corresponding eigenstates are represented in the Bloch sphere by the vectors
$\pm\hat{n}_{k}=\pm\left(\xi,0,\sin\varphi\sin k\right)/\sqrt{\rho(k)}$. 
For the lowest band [see Fig.\,\ref{BlochSpheres}(b),(c)], eigenstates with positive (negative) momentum, have positive (negative) spin $z$-component 
$\langle{\sigma_{z}}\rangle=\cos\theta_k$
, signalizing spin current separation [Fig.\,\ref{SpinMomentum}]. Time reversal invariance implies Kramers degeneracy for pairs of states $\hat{n}_{k}$ and $\hat{n}_{-k}$, which have opposite momentum and opposite 
$\langle{\sigma_{z}}\rangle$. The two modes composing a Kramers pair are:
\begin{eqnarray}
\cos\frac{\theta_{k}}{2}a^{\dagger}_{k}+\sin\frac{\theta_{k}}{2}b^{\dagger}_{k}, &&
\sin\frac{\theta_{k}}{2}a^{\dagger}_{-k}+\cos\frac{\theta_{k}}{2}b^{\dagger}_{-k}. 
\label{Kramers}
\end{eqnarray}
For the ladder system, spin-momentum locking implies that, on the average,  particles in the right leg move upwards, whereas particles in the left leg move downwards [Fig.\,\ref{SpinMomentum}], giving rise to a net chiral current that flows in the direction given by the magnetic flux. We characterize the chirality of the ladder modes by 
the quantity 
\begin{equation}
\mathcal{C}=k(\langle a^{\dagger}_{k}a^{}_{k}\rangle-\langle b^{\dagger}_{k}b^{}_{k}\rangle)=k\langle{{\sigma}_{z}}\rangle.
\label{Chirality}
\end{equation}
Within the lowest (highest) energy band, eigenstates have positive (negative) chirality. Within a band, the two states of a Kramers pair have the same chirality.

\begin{figure}[]
\includegraphics[width=\linewidth]{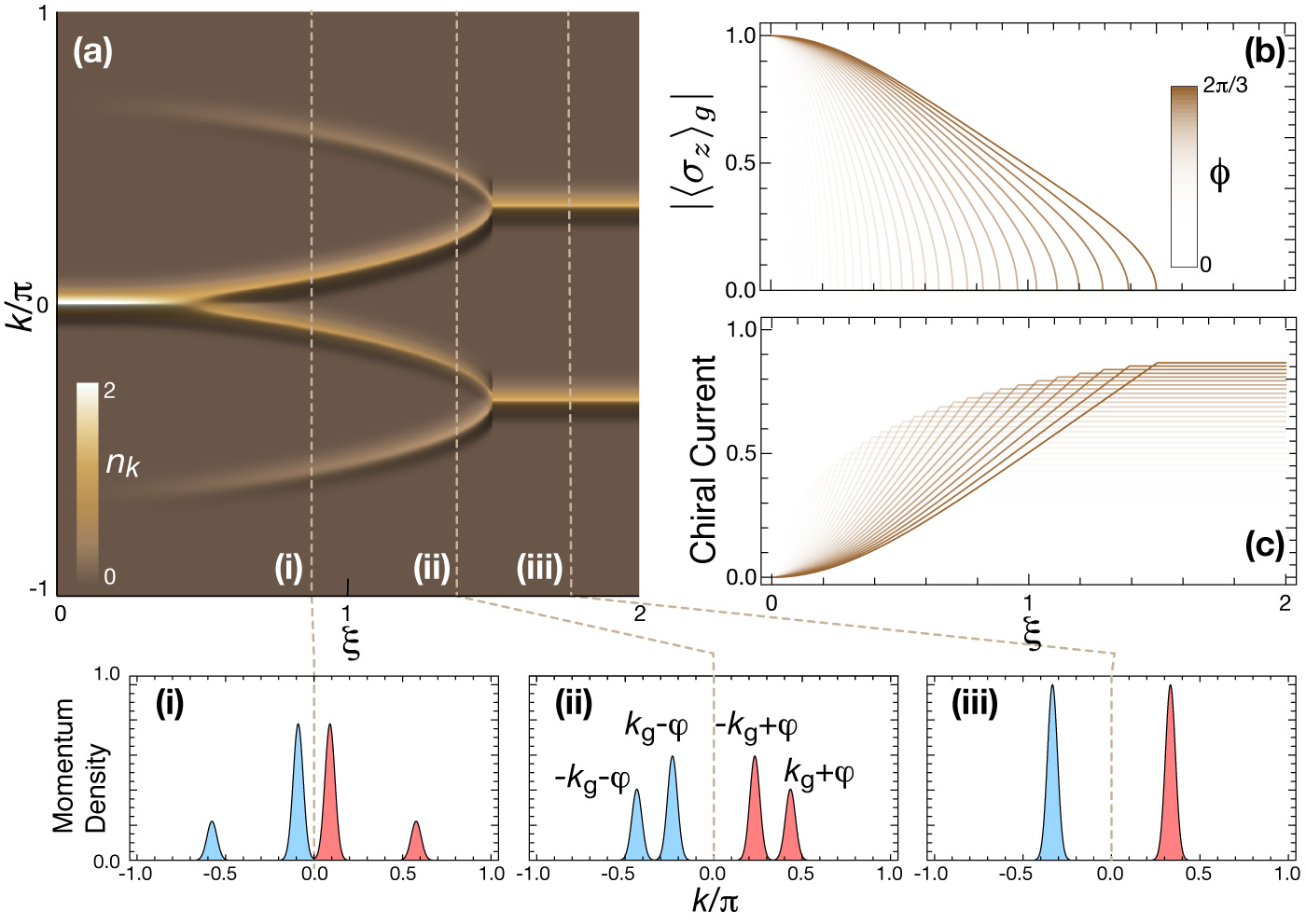}
\caption{\textbf{Time-of-flight observation of chirality and quantum phase transition}. 
({\bf a}) Ground state momentum density in the experimental gauge for increasing tunneling between legs, $\xi$, at flux $\phi=2\pi/3$. Below the transition point,  $\xi<\xi_c$ [({\bf i}),({\bf ii})]
four momentum peaks appear. They correspond to splitting of the two opposite momenta $\pm k_g$ of the two degenerate ground states in the theoretical gauge: $k_g\pm\varphi$, and $-k_g\pm\varphi$ (see text). For $\xi>\xi_c$, the two opposite momenta are merged into a single momentum component at $k=0$, which results into two peaks at $\pm \varphi$ for the experimental gauge. ({\bf b}) Chirality and ({\bf c}) chiral current of the ground state as a function of $\xi$ for different values of the flux $\phi$. At the critical point $\xi_c$, both magnitudes exhibit a discontinous derivative.}
\label{PhaseTransition}
\end{figure}

The lowest energy band exhibits two energy minima [see Fig.\,\ref{BlochSpheres}(a2)], characteristic of spin-orbit coupled systems, which correspond to two opposite spin-momentum states. Though the total physical current vanishes at these minima, the left and right current components are finite and opposite in magnitude, leading to a chiral current:
\begin{eqnarray}
J_\mathcal{C}&=&J_a-J_b=\langle {\sigma}_z \frac{\partial \mathcal{H}}{\partial k}\rangle,
\label{ChiralCurrent}
\end{eqnarray}
given by $J_\mathcal{C}=\cos^2\frac{\theta_{k}}{2}\sin(k-\varphi)-\sin^2\frac{\theta_{k}}{2}\sin(k+\varphi)$.
At a critical value of the tunneling between legs, 
$\xi_c=\sin^{2}\varphi/\cos\varphi$,
the two energy minima at $\pm k_g$ are merged into one single minimum at $k=0$ [see Fig.\,\ref{PhaseTransition}(a)]:
\begin{equation}
\sin{k_{g}}=\pm\sin\varphi\sqrt{1-\xi^{2}/\xi_c^2}.
\end{equation}
For bosonic particles, we predict a quantum phase transition at the bifurcation-merging point which is signalized by a discontinuous derivative of both the chirality $\mathcal{C}$ and the chiral current $J_\mathcal{C}$ [see Fig.\,\ref{PhaseTransition}(b),(c)].
This quantum phase transition is a consequence of the competition between the effective spin-orbit coupling and the effective magnetic field term in the ladder system.

\begin{figure}[]
\includegraphics[width=\linewidth]{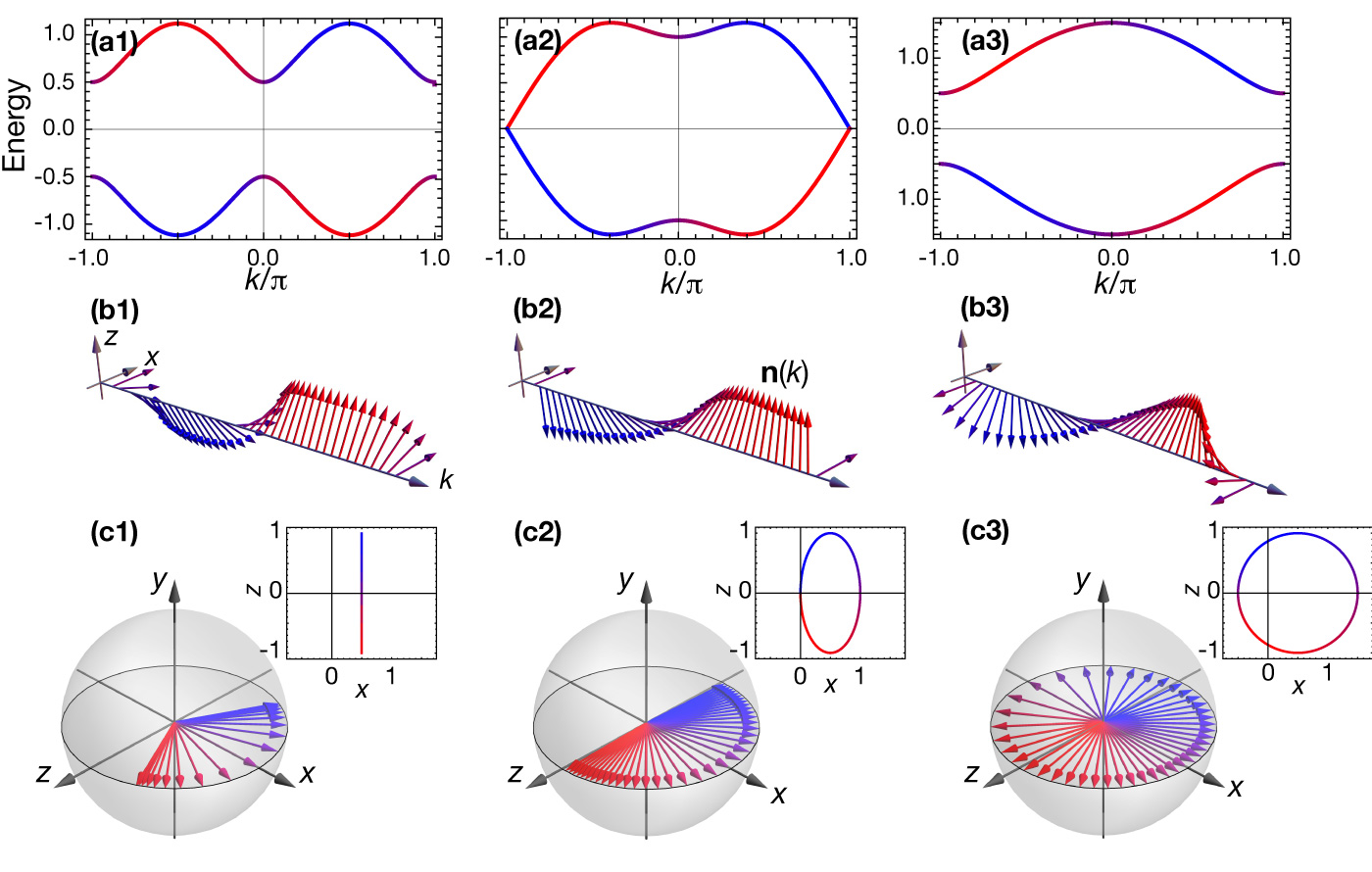}
\caption{\textbf{Topological phase transition with diagonal tunneling}. Ladder energy bands 
and lowest band eigenstates for flux $\phi=\pi/2$ and $\xi=0.5$, for different values of the diagonal tunneling ({\bf a1}) $\xi_d=0$, ({\bf a2}) $\xi_d=0.5$, and ({\bf a3}) $\xi_d=1$. At the critical point ({\bf a2},{\bf b2},{\bf c2}) the energy gap closes and a phase transition occurs from a trivial insulating phase ({\bf a1},{\bf b1},{\bf c1}) to a topological phase ({\bf a3},{\bf b3},{\bf c3}), where the spin texture characterizing the band winds once around the origin. The insets in {\bf c1},{\bf c2}, and {\bf c3} show the ellipse described by the tip of the vector $\rho(k)\hat{n}_k$, which encloses the origin above the transition point.}
\label{Diag}
\end{figure}

{\em Diagonal coupling and topological phase transition}. In the presence of diagonal tunneling, $t_d$, the ladder Hamiltonian takes the form:
\begin{equation} 
\mathcal{H}_d(k)=\mathcal{H}(k)+\xi_d \cos k \;{\sigma}_x,
\label{HamiltonianMomentumDiag}
\end{equation}
where $\xi_d=t_d/t$ and $\mathcal{H}(k)$ is the matrix Hamiltonian in Eq. (\ref{HamiltonianMomentum}). For $\xi_d/\xi>1$ the spin texture characterizing the bands depicts a closed loop in the $x-z$ plane of the Bloch sphere, and the system enters into a topological insulating phase [see Fig.\,\ref{Diag}]. At $\phi=\pi/2$
the ladder model in Eq.(\ref{HamiltonianMomentumDiag}) is equivalent to the Su-Schrieffer-Heeger (SSH) model \cite{SSH} describing polyacetylene, under the unitary transformation $U=(\sigma_z+\sigma_y)/\sqrt{2}$. 
By tuning the ladder parameters (for instance, by inducing a complex phase to the diagonal tunneling) the symmetry class of the topological phase can be changed, achieving the chiral unitary (AIII) class \cite{Schnyder}.


\begin{figure}[]
\includegraphics[width=\linewidth]{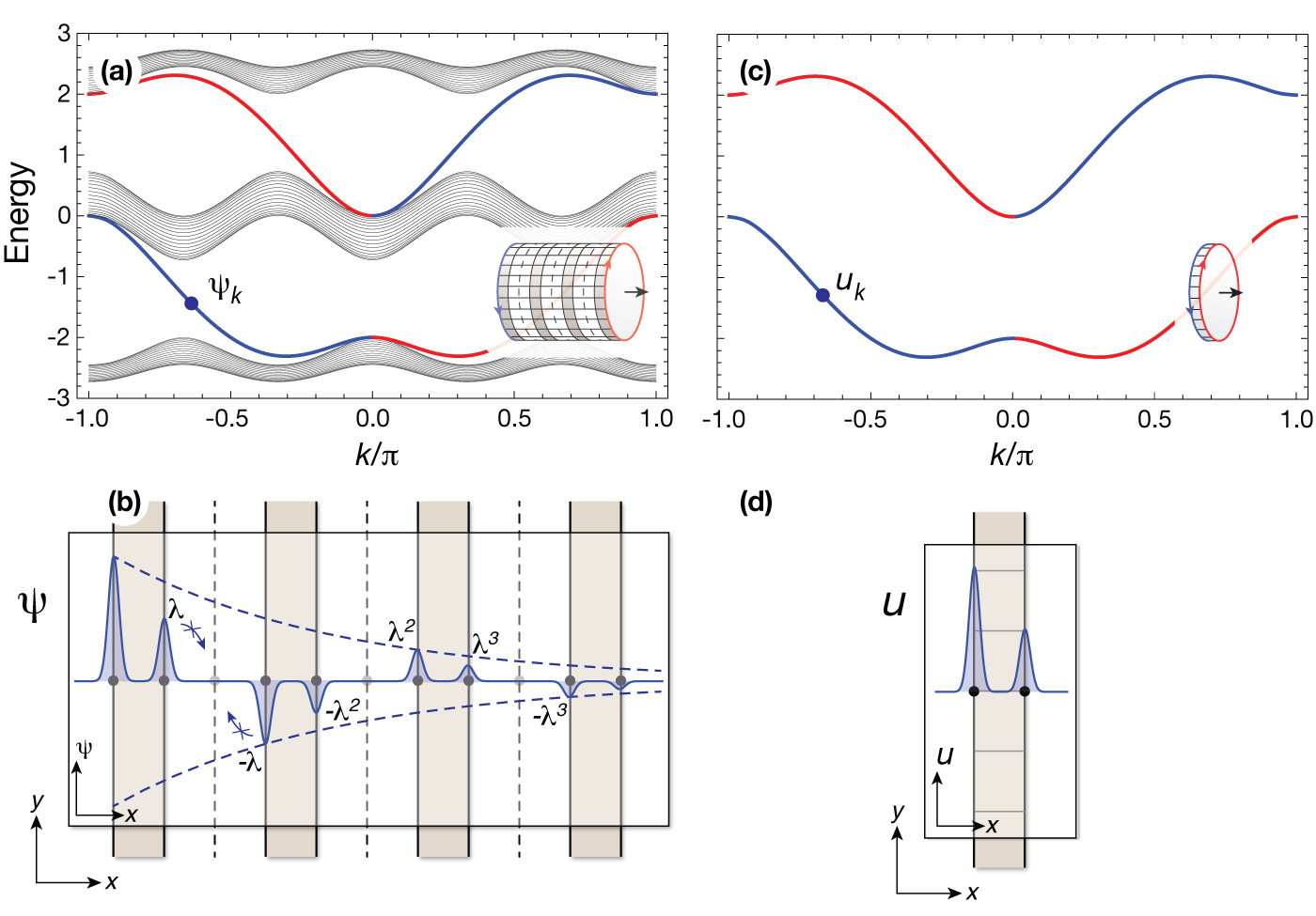}
\caption{\textbf{Chiral edges and ladder modes correspondence}. ({\bf a}) Energy bands for the Hofstadter model at $\phi=2\pi/3$ in a cylindrical geometry. The conducting chiral edge modes correspond to energies lying inside the gaps. They have the same energies as the modes of a two-leg ladder shown in ({\bf c}). Their spatial structure ({\bf b}) is made up of decoupled ladder modes ({\bf d}).
}
\label{HofstadterBand}
\end{figure}

{\em Ladder modes and chiral edge states of the Hofstadter model}.
We establish an exact correspondence between the chiral edge states of the Hofstadter model with flux $\phi=2\pi m/n$ per plaquette and the eigenstates of 
a ($n-1$)-leg ladder with the same flux and tunneling couplings.
The eigenenergies are identical. The edge states are found to be repetitions of ladder eigenstates, with an exponential decay.
Let us consider the Hofstadter model in a cylindrical geometry with rational flux $\phi=2\pi n/m$ per plaquette and tunneling couplings $t'$, $t$ along the $x$ and $y$ direction, respectively. Choosing a gauge that preserves translational invariance in the $y$-direction, the eigenstates  have well defined momentum $k$ in the $y$-direction and well defined  magnetic momentum $q$ in the $x$-direction. They are represented by $n$-component vectors 
$\Psi_{k,q}^{}$, which satisfy the set of Harper equations:
\begin{eqnarray}
	\label{HarperEquation}
\epsilon(k,q)\Psi_{k,q}^{(\alpha)}&=&\xi e^{-iq\delta_{\alpha,n}}\Psi_{k,q}^{(\alpha-1)}
+\xi e^{iq\delta_{\alpha,n}}\Psi_{k,q}^{(\alpha+1)}+\nonumber
\\&&\cos(k+\alpha\phi)\Psi_{k,q}^{(\alpha)}.
\end{eqnarray}
Here,  $\Psi_{k,q}^{(\alpha)}$ denotes the $\alpha$th eigenstate component, with $\alpha=1,\ldots, n$, and $\delta$ is the Kronecker delta.
The eigenstates form $n$ bands [see Fig\,\ref{HofstadterBand}(a)] characterized by non-trivial Chern numbers.

For a strip geometry, the Hofstadter model exhibits chiral edge modes which are localized at the boundaries of the 
cylinder [Fig\,\ref{HofstadterBand}(b)]. These modes  can be obtained from the equations above by substituting the magnetic momentum $q$ by $i\lambda$, where $\lambda$ is the localization length of the corresponding edge mode. The observation that we make in this work is that these edge modes, which we denote by $\psi_{k}$, correspond to the eigenmodes $u_{k}$ of an $(n-1)$-leg cut of the two-dimensional Hofstadter model. Direct substitution $q\to i\lambda$ in equation (\ref{HarperEquation}) shows that they satisfy:
\begin{eqnarray}
\psi_{k}^{(\alpha)}=
\begin{cases}
u_{k}^{(\alpha)}, & \text{with} \;\; \mathcal{H}_{n-1}(k)u_{k}=E_{n-1}u_{k} 
\\ &\text{if } \alpha=1,\ldots,n-1
\\
0 & \text{if } \alpha=n,
\end{cases}
\end{eqnarray}
where  $\mathcal{H}_{n-1}(k)$ is the Hamiltonian of the $(n-1)$-leg ladder in momentum space. 
These edge states are therefore repetitions of decoupled ladder modes in the form 
$e^\dagger_{k} \propto \sum_j \lambda ^{j} d^\dagger_{k,j}$,
where 
$d^\dagger_{k,j}=\sum_{\alpha} u_k^{(\alpha)} a^\dagger_{k,nj+\alpha}$ is an eigenmode of the $j$th ladder.
The localization length is given by $\lambda(k)=-u_k^{(n-1)}/u_k^{(1)}$, guaranteeing that the amplitudes at the end and the beginning of neighbouring ladders are opposite, so that tunneling to sites between ladders vanishes [see Fig\ref{HofstadterBand}(b)].
For $\phi=2\pi/3$, the chiral edge modes correspond to the chiral modes of the two-leg ladder we analyzed above [see Fig\,\ref{HofstadterBand}(c),(d)]. The localization length is given by
$\tanh \lambda(k)=\langle\sigma_z\rangle$. It is interesting to see how the chirality of the ladder modes is directly connected to the chiral and localized character of the modes of the two-dimensional system. 



{\em Experimental observation}. 
The ladder properties we describe above can be readily observed in an actual experiment by combining different techniques available in current optical lattice-based experiments.
An array of decoupled ladders with magnetic flux could be realized by using a superlattice structure together with the recently created staggered flux patterns in a two-dimensional optical lattice \cite{Aidelsburger2011}.
We propose the following schemes to observe spin-momentum locking and the corresponding chirality of ladder modes. i) {\em State preparation}. By inducing Bloch oscillations in the lowest band, particles can be prepared in any desired momentum eigenstate \cite{Atala2012}. ii) 
{\em Momentum peaks and chirality}. If particles are released from the optical lattice, the momentum components of the corresponding momentum eigenstate can be revealed in a time-of-flight measurement. For the gauge in which experiments are typically performed \cite{Aidelsburger2011}, eigenstates are obtained from the ones discussed above by the following gauge transformation:
\begin{equation}
a^{\dagger}_{k}\:\rightarrow\:a^{\dagger}_{k-\varphi},\;\;\;
b^{\dagger}_{k}\:\rightarrow\:b^{\dagger}_{k+\varphi}.
\end{equation}
This implies that the two degenerate states within a Kramer pair (Eq.\,\ref{Kramers}) are transformed into
\begin{eqnarray}
\label{formelab}
\cos\frac{\theta_{k}}{2}a^{\dagger}_{k-\varphi}+\sin\frac{\theta_{k}}{2}b^{\dagger}_{k+\varphi}, &&
\sin\frac{\theta_{k}}{2}a^{\dagger}_{-k-\varphi}+\cos\frac{\theta_{k}}{2}b^{\dagger}_{-k+\varphi}. \nonumber 
\end{eqnarray}
Thus four momentum components at $k_g\pm\varphi$ and $-k_g\pm\varphi$ should be observed 
[see Fig.\,\ref{PhaseTransition}(a)].
Independent detection of even and odd legs would reveal the different heigths and positions of left and right momentum components, directly reflecting spin-momentum separation. Combining the information about momenta position and height, the chirality of the mode can be obtained.
iii){\em Phase transition for bosonic particles}. By modyfing the superlattice potential, tunneling between ladder legs can be tuned, driving a quantum phase transition. A time-of-flight experiment will show merging of the four momentum peaks into two peaks at the transition point [see Fig.\,\ref{PhaseTransition}(a)]. The transition also becomes manifest in the behaviour of  the chiral current of the ground state, which could be directly measured by suddenly splitting the system into decoupled vertical bonds. Within each double well the system will then perform density oscillations with an  amplitude equal to the initial current along the bond. Bonds in even arrays should be anticorrelated with those in odd arrays, reflecting the non-vanishing chirality of the ground state. At the critical point, $\xi=\xi_c$, the oscillation amplitude should show a kink [see Fig.\,\ref{PhaseTransition}(c)], indicating the transition.

The correspondence between ladder modes and the chiral edges of the Hofstadter model opens up a new possibility for the detection of the latter. A cylindrical geometry could be achieved by connecting the two edges along one direction of a two-dimensional lattice. Topological edge states could then be probed by converting the system into an array of decoupled two-leg ladders using a superlattice structure. As a direct manifestation of the key features characterizing the chiral edges modes of the Chern insulator, we should observe that i) time-of-flight imaging remains invariant before and after decoupling into  ladders, and ii) independent detection of left and right ladder legs show spin-momentum locking, exactly as for the ladder discussed above. 

{\em Conclusion and outlook}. We have shown chiral ladders to be promising link systems between one-dimensional 
spin-orbit (time reversal) insulators and two-dimensional topological Chern insulators. 
On the one hand, they provide us with a realization of spin-orbit interaction in atomic systems which is not hindered by 
undesired light scattering processes. On the other hand, they are quantum simulators of the physics at the chiral edges of Chern insulators, exactly reproducing their chiral, localized and robust character. We have developed realistic schemes in which the topological and chiral properties of the ladder system could be readily probed in optical lattice-based experiments.
Furthermore, we expect a rich interplay between particle interactions in the ladder and the topological and chiral features we describe in this work, possibly giving rise to interesting topological phases, such as fractional topological insulators and topological superfluids. In this direction, the ladder-built realization of spin-orbit coupling opens an interesting path towards the exploration of a topological atomic superfluid. By inducing s-wave pairing in the left-right leg degree of freedom, a $p$-wave superfluid state could be achieved via bringing the $s$-wave superfluid into contact with the spin-orbit insulator. It is a challenge to develop schemes to realize such a $p$-wave superfluid and to probe and manipulate the emerging Majorana fermions.

\appendix


\end{document}